\documentclass[preprint,authoryear,12pt]{elsarticle}

\usepackage{graphicx}

\usepackage{amssymb}

\usepackage{lineno}

\journal{Journal of Theoretical Biology}

\begin{document}

\begin{frontmatter}



\title{On Species Persistence-Time Distributions}


\author[1]{S. Suweis}
\author[1]{E. Bertuzzo}
\author[1]{L. Mari}
\author[2]{I. Rodriguez-Iturbe}
\author[3]{A. Maritan}
\author[1,4]{A. Rinaldo}
\address[1]{Laboratory of Ecohydrology (ECHO/IIE/ENAC), Ecole Polytechnique F\'ed\'erale  Lausanne (EPFL), School of Architecture, Civil and Environmental Engineering (ENAC), 1015 Lausanne, (CH)}
\address[2]{Department of Civil and Environmental Engineering, Princeton University, Princeton, NJ 08544, (USA)}
\address[3]{Department of Physics, University of Padua, CNISM and INFN, via Marzolo 9, 35131 Padova (Italy)}
\address[4]{Dipartimento IMAGE, University of Padua, via Loredan 20, I-35131 Padova (Italy)}

\begin{abstract}
We present new theoretical and empirical results on
the probability distributions of species persistence times in natural ecosystems.
Persistence times, defined as the timespans occurring between species' colonization and local extinction in a given geographic region, are empirically estimated from local observations of species' presence/absence. A connected sampling problem is presented, generalized and solved analytically. Species persistence is shown to provide a direct connection with key spatial macroecological patterns like species-area and endemics-area relationships. Our empirical analysis pertains to two different ecosystems and taxa: a herbaceous plant community and a estuarine fish database. Despite the substantial differences in ecological interactions and spatial scales, we confirm earlier evidence on the general properties of the scaling of persistence times, including the predicted effects of the structure of the spatial interaction network. The framework tested here allows to investigate directly nature and extent of spatial effects in the context of ecosystem dynamics. The notable coherence between spatial and temporal macroecological patterns, theoretically derived and empirically verified, is suggested to underlie general features of the dynamic evolution of ecosystems.
\end{abstract}


\end{frontmatter}


\section{Introduction}
\label{I}
Extinction rates, pointedly from habitat loss, have been defined as a crucial conservation problem of this century \citep{Raup1984,Wilcox1985,Pimm1988,Fahrig1997,Ferraz2003}. Direct and reliable estimation of true extinction rates remains problematic, however, and a lively debate exists over the possible overestimation of extinctions incurred through the standard use of methods based on the reversing of the species-area relationship (SAR), i.e. the number of species observed at increasing sampled areas \citep{green2003,he2011}.

Recently \citet{bertuzzo2011} have proposed empirical and theoretical evidence for a new macroecological pattern, namely the distribution of persistence times for trophically equivalent, co-occurring species. The species persistence time (SPT) is defined as the time-span incurred between a species' emergence and its local extinction in a given geographic region. Using two different long-term ecological databases, related namely to the inventory of North American breeding birds and to Kansas grasslands, \citet{bertuzzo2011} suggested that the empirical SPT distribution is characterized by a power-law function
limited by an exponential cut-off determining the maximum observed persistence time, which in turn is related to the spatial extent of the ecosystem.  The advantage to concentrate on SPTs, is that SPT distributions have proved to be a robust measure of species turnover at different spatial scale. Although SPT distributions cannot provide predictions about extinction rates of specific species, they can describe the global evolution of the diversity of an ecosystem and give robust estimates for mean SPTs \citep{bertuzzo2011}.

The problem of the scale of observation becomes central in this framework. In fact, at the local scale, say an observational site for the presence/absence of breeding birds, the persistence time of a species is controlled by ecological processes operating at short timescales, like birth/death, dispersal, and
contraction/expansion of geographic ranges. At this scale local extinctions are dynamically balanced by colonizations
\citep{machartur1967,tilman1994,Ricklefs2003,Muneepeerakul2008}. At the global scale,
species emergence and extinction are controlled by mechanisms acting
on evolutionary timescales \citep{brown1977, Diamond1989}.
Interestingly, the scaling behavior proposed by \citet{bertuzzo2011} to govern the transition from local to global scales is capable of effectively describing the overall dynamical evolution of the
ecosystem diversity. Scaling features also allow to predict SPT distributions for wide geographic areas from measures of persistence on smaller areas, which are much easier to monitor.

To provide a theoretical explanation for the SPT
pattern, \citet{bertuzzo2011} resorted to a
spatially explicit model based on neutral theory.  Neutral theory has been proposed as a
unifying theoretical context for understanding ecological patterns
\citep{UNTB}. Since its formulation, several studies have focused
their attention on it
\citep[e.g.][]{volkov2003,Houchmandzadeh2003,chave2004,Alonso2004,azaele2006,Muneepeerakul2008,chilsom2009}.
Neutral models are based on the assumption that, within a single
trophic level, individual birth and death rates are
species-independent. The main advantages of neutral models are that they are falsifiable, and that they are able to generate predictions for many different macroecological patterns. This is the
case, for instance, of the relative species abundance (RSA)
distribution \citep{McGill2003,volkov2005}, the
species-independent beta diversity patterns
\citep{condit2002,Zillio2005}, species geographic range
\citep{Bertuzzo2009} and the species-area relationship
\citep{brown1995,zillio2008,ODwyer2010}. Remarkably, this is
accomplished by invoking only basic ecological processes such as
birth, death, migration and dispersal limitation. Although certain
emergent ecological patterns are independent of the fine
ecological details and often well predicted by neutral theory
\citep{Bell2001}, this does not imply that the underlying
ecological processes are neutral \citep{harte2003,Pacala2005}.
Although the SPT distribution framework does not necessarily require
neutral processes as well, we shall keep them as a reference
modelling frame for our theoretical speculations.

This paper explores the subject of the SPT distribution further by
extending the empirical and theoretical analysis presented in
\cite{bertuzzo2011}. In particular, we report here on the study of
two new long-term datasets to test the robustness of the SPT
macroecological pattern. Moreover, we study in details the
connections of the SPT distribution with the structure of the SAR
and the endemics-area relationship \citep[EAR, which gives the number of
species that are confined within the sampled area;  see][]{Kinzig2000}.  We also investigate the role of spatial
interactions in the new SPT data and show that the effect of the
environmental matrix \citep{Ricketts2001} on SPT distributions
depends on the overall geographic area in which the ecosystem is
embedded: if the typical dispersal range is comparable to the
total area, possibly because of dispersal limitation, then spatial
interactions may be neglected. Finally we present and solve
analytically a new particular case of a length-biased sampling
problem \citep{Cox1969} that arises in SPT data analysis; namely
we find how to relate the SPTs derived from finite observational
windows to the inherent SPT distribution of the ecosystem.

The paper is organized as follows: in the next section we briefly
summarize the spatially explicit model for SPTs. In subsection 2.1 we
discuss the effects of data sampling on SPT distributions. We
present a general mathematical framework to relate SPT theoretical
distributions to those obtained from finite samples. In the third
Section, we present novel empirical SPT
distributions regarding two different ecosystems (a forest in New
Jersey and an assembly of estuarine fish in the Bristol Channel) and compare them
with the distributions predicted by the neutral theory. The role of
spatial scales and their interactions with SPT distributions, and
the intimate relation between the SPT distribution and the SAR/EAR
are also shown. A set of conclusions closes the article.

\section{The Model}

Neutral theory \citep{UNTB}, which assumes that all the individuals within
a given trophic level are competitively equivalent, offers a benchmark dynamics suggesting that
many aspects of real biotic patterns may not require a more complex model \citep{Muneepeerakul2008,Bertuzzo2009}.
According to the assumption of neutrality \citep{UNTB}, the
dynamics of a species in the ecosystem is fully specified by its
effective birth and death rates $b(n)$ and $d(n)$, which depend
exclusively on the population size $n$. Species abundance dynamics
is described by the so called birth-death master equation (ME)
\citep{volkov2003,Houchmandzadeh2003}:
\begin{equation}\label{bdME}
    \frac{dP(n,t)}{dt}=b(n-1)P(n-1,t)+d(n+1)P(n+1,t)-(b(n)+d(n))P(n,t),
\end{equation}
where $P(n,t)$ is the probability, for a given species, of having a population of $n$ individuals at time $t$.
We note that $b(n)$ and $d(n)$ take into account several ecological processes
that may increase or decrease the number of individuals in a species over time as, for instance, immigration or emigration.
Assuming absorbing boundary conditions in $n=0$, the probability density function (pdf) of the theoretical persistence time $p_{\tau}$ can be expressed as \citep{pigolotti2005}:
\begin{equation}\label{pt_pdf}
    p_{\tau}(t)=\frac{d P(0,t)}{dt}.
\end{equation}
Considering $b(n)=n(1-\nu)$ and $d(n)\propto\,n$, the above model
describes analytically the mean field dynamics (in the limit of
infinite number of nodes) of the following lattice model, known as
the voter model \citep{Liggett1975,Durrett2002,Zillio2005}. $\nu$
is termed diversification rate and it measures the frequency of
appearance of new species in the system. Let us consider, for
simplicity, a system with $N$ nodes
($N\rightarrow\infty$). Each node may be occupied by a single
individual belonging to a species which is represented by a given
color. The dynamics at each time step is as follows. A randomly
selected individual dies. With probability $\nu$ this site is
occupied by a new species; with probability $1-\nu$, the site is
colonized by an offspring of a node randomly chosen in the lattice
(mean field case). We note that in this framework we
require that the probability that a new species enters the system
be uniformly distributed in time. This seems like a reasonable
assumption in the absence of strong environmental or historical
gradients within the considered timescale. 

The spatial version of this model allows colonizations to occur
only from one of the neighbors of the empty patch (i.e. dispersal
limitation). The same model can be applied to 1D/3D lattices, or
to a networked environment where dispersal is constrained
directionally. By tracking every individual up to a statistically
stationary state, it is possible to build the time series of
species abundances and, from it, to compute numerically the SPT
distribution. SPTs  prove
crucially dependent on the type of spatial connectivity
interactions in the voter model. Exact and numerical results for
the spatial voter model \citep{bertuzzo2011}, in fact, show that
the SPT distribution is consistent with power-law shapes of the
type:
\begin{equation}\label{SPT_distribution}
    p_{\tau}(t)=\mathcal{C} t^{-\alpha}e^{-\nu t},
\end{equation}
where the scaling exponent $\alpha$ strictly depends on
the connectivity structure and $\mathcal{C}=\nu ^{1-\alpha}/\Gamma
(1-\alpha)$ is the normalization constant ($\Gamma(x)$ is the
complete Gamma function of argument $x$ ). The neutral model thus predicts the same functional shape for the STP distribution, a power law with the exponent $\alpha$ that is solely determined on the connectivity structure of the environment (Figure 1). Therefore $\alpha$ may be seen as a critical exponent  \citep{StanleyHE} that is, it  describes the behavior of physical systems near phase transitions. In general it is argued that scaling exponents  do not depend on the details of the physical system, but only on the dimension $D$ of the system (e.g.
$D=1$ on a line, $D=2$ on an isotropic lattice and so on) or the
range of the interactions \citep{StanleyHE}. This is not the first
time that similarities between ecosystem dynamics and critical
systems in physics have been pointed out
\citep{finitesize,zillio2008}. 

From Monte Carlo simulations of the spatial voter model, $\alpha$
exponents for different geometric cases can be evaluated. It has
been shown \citep{bertuzzo2011} that $\alpha\approx1.5$ for the
1D lattice, $\alpha\approx1.62$ for the networked landscape,
$\alpha\approx1.82$ and $\alpha\approx1.92$ respectively for the
2D and 3D lattice (Figure 1). In the mean field
case, the scaling limit ($\nu \ll 1$) of $p_{\tau}(t)$ can be
calculated exactly \citep{pigolotti2005} yielding $
p_{\tau}(t)\sim  t^{-2}f(\nu t)$, with $f(z)=(z/(1-e^{-z}))^2
e^{-z}$. The robustness of the scaling behavior as a function only
of the structure of the environmental matrix of interactions is
deemed remarkable.

\subsection{An Ecological Length-Biased  Sampling Problem}

Empirical data on persistence times for different species are
available, especially from succession studies and long-term
ecological databases \citep{pigolotti2005,bertuzzo2011}. However,
for further empirical analysis, sampling effects on SPTs need to
be taken into account. This type of sampling problem belongs to
the class of so called length-biased problems, that arise in a
wide range of different disciplines \citep{patil1978,vardi1982}.

We start with a thought experiment which we refer to as the
stick-length sample problem. Consider an infinite number of sticks
of different random length $L$ distributed according to
some unbiased pdf, say $p_{L}$. Imagine to place randomly
these sticks on a plane $x$-$y$, aligned in the same direction
(e.g. along the $x$ direction). The process is thus
defined in a 1D space, and the $y$ axis is just a way to visualize
different realizations of the same process.
We denote by $X^0$ the position of the left side of
a given stick. $X^0$ is a random variable uniformly
distributed in $(0,\infty)$. Consider only those sticks that
cross a vertical line $x=x^*$ and measure the lengths of those
sticks (Figure 2).

We will denote by $p_{L_c}$ the length
distribution of this sticks sample. The stick-length sample
problem can be formulated with the following questions: a) how
is $p_{L_c}$ related to $p_{L}$? b) How are the left and right ends of
the sticks that cross $x^*$ (indicated by $L_{p}$, Figure 2)
distributed? We now show how length measurements on particular
samples of the stick population give length distributions which
are different from that of the whole stick population.

Using tools of probability theory, it can be shown (Appendix A)
that the pdf of the length of the sticks that cross $x^*$, in the
limit $x^*\rightarrow\infty$ (to avoid boundary effects), is
\begin{equation}\label{p_L_c}
    p_{L_c}(l)=\frac{l p_{L}(l)}{\langle L \rangle},
\end{equation}
where $\langle L \rangle = \int_0^{+\infty}L p_{L}(L)dL$ is the
average stick length. Thus the length pdf of the sample
constituted by those sticks that cross the observation point
$x^*$, $p_{L_c}$, is not the same as the overall population stick
length pdf, i.e. $p_L$.

Let us now introduce a second vertical line $x=\hat{x}<x^*$. We now want to 
measure the length of the portions of sticks $L_p$ crossed by the two vertical lines ( $x=\hat{x}$ and $x=x^*$) 
that are comprised inside the space
window $\Delta x=x^*-\hat{x}$ (see Figure 2). If we
denote by $ p_{L_p}$ the pdf of $L_p$, then it can be shown
(Appendix A) that:
\begin{equation}\label{p_L_m_sol}
    p_{L_p}(l)=\frac{\int_{l}^{\infty}p_{L}(L)dL}{\langle L \rangle}.
\end{equation}

Finally, we can calculate the pdf $p_{L'}$ of the random
variable $L'$ describing the length of a stick that is totally or
partially comprised within the sampling window $\Delta x$. We
note that in this case the maximum measured length of a
stick is $\max\{L'\}=\Delta x$. $L'$ can be expressed as a
function of the other independent random variables $L$ and
$X^0$, that have already been probabilistically characterized. in
fact, we can distinguish four different cases (see Figure 3):
\begin{itemize}
\item the stick is completely inside the window, then $L'=L$;
\item $X^0>\hat{x}$ and the stick crosses $x=x*$, then $L'=x^*-X^0=L_{p}$
\item $X^0<\hat{x}<L+X^0<x^*$ and the stick crosses $x=\hat{x}$, then $L'=X^0+L-\hat{x}=L_{p}$;
\item the stick is longer that the space window, then $L'=\Delta x$.
\end{itemize}
As explained in detail in Appendix A, we find that:
\begin{eqnarray}
p_{L'}(l)&=&\frac{1}{\mathcal{N}}\bigg((\Delta x-l)p_{L}(l)\Theta(\Delta x-l)\;+\nonumber\\
 &+& \Theta(\Delta x-l)\int_{l>0}^{\infty}p_{L}(L)dL\;+\nonumber \\
&+& \Theta(\Delta x-l)\int_{l>0}^{\infty}p_{L}(L)dL\;+\nonumber \\
&+& \delta(l-\Delta x)\int_{\Delta x}^{\infty}(l-\Delta x)p_{L}(L)dL \bigg),\; \label{pL_final}
\end{eqnarray}
where $\mathcal{N}$ is the normalization constant for $p_{L'}$, and
$\delta(x)$ and $\Theta(x)$ are the Dirac delta distribution and
the Heaviside step function, respectively. The last term of
equation (\ref{pL_final}) gives an atom probability in $L=\Delta
x$ corresponding to the fraction of sticks longer than the total
length of the observational window ($\max\{L'\}=\Delta x$).

The probability distribution $p_{L_I}$ of the sticks'
length inside the window $\Delta x$ follows directly from the
first term of equation (\ref{pL_final}):
\begin{equation}\label{pL_i}
p_{L_I}(l)\;=\;\frac{1}{\mathcal{N'}}(\Delta
x-l)p_{L}(L)\Theta(\Delta x-l),
\end{equation}
where $\mathcal{N'}$ is a proper normalization constant.

The analytical results provided by Eqs. (\ref{pL_final}) and (\ref{pL_i}) are tested via numerical simulation, as shown in the bottom panel of Figure 2. First we generate $10^5$ segments along the $x-axis$ of random length drawn from an exponential pdf, i.e. $p_L(l)=\lambda e^{-\lambda l}$. Afterwards, we consider only those segments that are within the spatial window $\Delta x=1/\lambda$ and we reconstruct the stick length pdf  $p_{L_I}$. Analogously, we build $p_{L'}$. Finally we compare the numerical distributions with the the solution of Eqs. (\ref{pL_final}) and (\ref{pL_i}). As expected the agreement is perfect.

\subsubsection{Application to SPTs.}

Several real ecological sampling problems, both in time and space,
can be mapped into the thought experiment described in the
previous Section. In particular we now apply these results to
SPTs.

Assume that species abundance (or presence-absence) time series
for a given ecosystem characterized by  a single trophic level are
available from field campaigns carried on in a time period of span
$\Delta t_w=t_f-t_0$ years. From the collected data, SPTs can be
measured and the empirical SPT distribution likewise obtained.
This distribution does not correspond to the theoretical SPT
distribution, $p_{\tau}$, in the same way as $p_{L}\neq p_{L'}$ in
the stick sample problem. In particular, $p_{\tau}(t)$ is affected
by the fact that species persistence can be recorded only within a
finite temporal window $\Delta t_w$.  Thus, given the theoretical
persistence-time pdf described by Eq. (\ref{SPT_distribution}), we
are interested in the pdfs of the variables that we can actually
measure, i.e. the persistence times $\tau'$ of those species
emerging before or during the observations and that are still
present or go locally extinct within the time window $\Delta t_w$
(and the persistence time $\tau_I$ of only those species that
emerge and go locally extinct within the observed time window).

In our theoretical framework, the probability of a diversification
event in a time step is constant. Thus the emergences of new
species in the system obey a Poisson process, say $U(t)$, with
rate $\lambda=\nu\,N$, where $N$ is the total number of
individuals in the system and $\lambda$ has the dimensions of the
inverse of (generation) time. Consequently the emergence times
$T^0$ of species in the system are uniformly distributed.
Therefore, we can map exactly this situation  into the previous
stick-length sample problem, with SPTs replacing stick lengths,
i.e. $L\rightarrow\tau$, $L'\rightarrow\tau'$,
$L_I\rightarrow\tau_I$ and $X^0\rightarrow T^0$. Eqs.
(\ref{pL_final}) and (\ref{pL_i}) allow us to test the theoretical
SPT pdf $p_{\tau}(t)$ obtained by the spatial model against the
empirical SPT distribution obtained from the measurements.

It is interesting to note how our thought experiment uses
similar approaches to the model known as the mid-domain effect, which describes 
how species� ranges that are randomly shuffled within a bounded
geographical domain overlap increasingly toward the center of the domain, creating a �mid-domain�
peak of species richness \citep{Colwell2004,arita2004}. In fact, SPTs are essentially
distributions of species temporal ranges, and therefore, the
sampling problem just presented may be interpreted as a sort of
mid-domain problem in time. However there is a crucial difference
between the two effects: while for spatial ranges there are
physical geometric constraints imposed by hard boundaries in space
(i.e. geographic boundaries), and thus species spatial ranges must
be contained within the given geographic region
\citep{Colwell2004,arita2004}, temporal ranges do not have constraints  in time. It is in fact to be reminded that 
species temporal ranges are not
limited by the observational time window, rather is our capacity to
measure real  species temporal ranges that is limited.

\section{Results.}

\subsection{Empirical SPTs and Comparison with Model Results.}

\cite{bertuzzo2011} presented a comparison between the theoretical
model and the empirical SPT distributions for two different
ecosystems (North American breeding birds and Kansas grassland)
which is shown in Figure 4a,b for purpose of completeness.

We now provide new evidence of the existence and
robustness of the SPT distribution pattern by presenting
empirical data on persistence times taken from two different
ecological databases and comparing them with the analytical
results on SPT distributions. Specifically, we empirically
characterize SPT distributions of two long-term datasets
concerning ecosystems with very different characteristics: 1) a
44-year long study from the Buell-Small Succession (BSS) Study  of
plants in New Jersey ; and 2) a 28-year long database of British
estuarine fish collected at Hinkley Point (headland on the Bristol
Channel coast of Somerset, England).

The BSS study \citep{BSS} includes ten fields that were released
from agriculture and used to study succession dynamics. Permanent
plots, measuring 48 m$^2$ each, were established in every field.
The permanent plots were sampled every year from 1958 to 1979,
after which they were sampled alternatively every other year to
the present day. The fields differ in the year of release, thus we
consider only data collected after the latest field abandonment
year (1968). In order to avoid missing data in alternate years,
the minimum sampled area (afterwards named cell) considered in the
calculation of the empirical SPT distributions comprises two
adjacent fields (96 m$^2$). In this way each cell is populated
with presence-absence records for each year. We repeat the same
analysis for 3 other different scales: 4,6,8 adjacent fields
(A=192, 288, 384 m$^2$, respectively). We also analyze the whole
system, which corresponds to an area of $A=A^*= 480$ m$^2$. For
every scale $A$ of analysis, the presented results refer to the
average over all possible combinations of adjacent plots within
the system (moving average procedure). A three-dimensional
presence-absence matrix $P$ is in this way built. Each element
$P_{stc}$ of the matrix is equal to 1 if species $s$ is observed
during year $t$ in a cell $c$, otherwise $P_{stc}$ = 0. The
empirical SPT is defined as the number of consecutive years in
which the measurements reveal the presence of the species in that
geographic region (see Figure 3). For every cell $c$ and every
species $s$ we measure persistence times from presence-absence
time series derived from the second dimension/index of matrix $P$.
The presence-absence time series thus form a vector of length $T$,
where $T$ is the total number of years of observation, that has as
$j$-th component a one if species $s$ is observed in cell $c$
during the $j$-th year. Persistence time is defined as the length
of a contiguous sequence of ones in the time series (Figure 3).
For every scale of analysis we consider all the measured
persistence times regardless of the species they belong to, and we
build the SPT empirical pdf at every cell scale (Figure 4c).

As for the second database, fish samples were collected from the
cooling-water filter screens at Hinkley Point �B� Power Station,
situated on the southern bank of the Bristol Channel in Somerset (England) from 1980 to 2008.
A full description of the intake configuration and sampling methodology  is given in
\citet{magurran2003}. We only consider estuarine fish species (not
crustaceous organisms). Empirical SPT pdf's can be computed as
described for the BSS dataset (Figure 4d). Note, however, that in
this dataset there is no spatial scale variability, because the
samples were collected in a single point in space. Therefore we
limit our analysis to a single spatial scale.

Using  Eqs. (\ref{pL_final}) and (\ref{pL_i}) and the
observational data we performed the best fit of $\alpha$ and
$\nu$, the parameters of the theoretical SPT pdf.  At any spatial
scale, the scaling exponent and the diversification rate for the
empirical SPT distributions are determined with a simultaneous
nonlinear fit of observational and analytical pdfs of $\tau'$ and
$\tau_I$. Table 1 summarizes the results of the fit. Remarkably,
in both cases the scaling exponents derived empirically  are
consistent with those predicted by the neutral voter model. In
particular, for the fishes we find $\alpha=1.97\pm0.06$, which is
compatible with the exponent $\alpha=2$ predicted by the mean
field voter model (see Figure 4d). In fact, no spatial information
is embedded in our empirical data and thus we do not expect to see
the effect of dispersal limitation on the relevant ecosystem
dynamics. Figure 4c shows the case of the New Jersey plant
communities, for which the best fit of  $p_{\tau}$ is
$\bar{\alpha}=1.97\pm0.12$ (where $\bar{\alpha}$ is the average of
the exponents over all the cell scales), which is also compatible
with the mean field prediction $\alpha=2$. In this case the
negligible effects of dispersal limitation on the ecosystem
dynamic suggests that the average plant dispersal radius $R$ is
comparable with the linear dimension of the total sampled area,
i.e. $R\approx \sqrt{A^*}$, and thus the net result is a mean
field dynamics not affected by spatial interactions
\citep{bertuzzo2011}.  The fact that the total sampled area of the
ecosystem is very small ($A^*=480$ m$^2$) strongly supports this
interpretation of the data.

While studying SPTs in a given
ecosystem on the basis of presence/absence (or count) data,
imperfect detection of species is a source of concern \citep{nichols98a}.
Species, in particular animal species, are, in fact, routinely sampled with a detection probability much smaller than one.
Although in general the problem of imperfect detection is somewhat less relevant
for herbaceous plant datasets, in the case of the BSS forest sampling occurs either annually or biannually, e.g. implying that seasonal variability in plant composition may be not picked up (and this varies from year to year).
However, plants that go undetected because of yearly sampling affect the estimates of persistence times shorter than one year, thus not an issue for the scaling of SPTs. Moreover, plants in vegetative state are not identifiable, nor are small recruits, typically.  For the fish database the power station intakes at
Hinkley Point are an effective sampler because of their location
at the edge of a large inter-tidal mudflat in an estuary. The
filter screens have a solid square mesh of $10$ mm and start to
retain fish of standard length of about $25$ mm \citep{Henderson1991}. A
99\% retention for many species occurs at the standard length of $40$ mm.
Nevertheless, incomplete sampling is present because only a small fraction of the water in the
Bristol channel is filtered, and  filtering thus only catches fish that cannot avoid being sucked by the water turbines.  Hence, even  in intensive sampling programs is difficult to tell whether absences are real or caused by incomplete detection/sampling. Note that the impact of imperfect sampling on SPTs of North America breeding birds has been analyzed in details in \citep{bertuzzo2011}. However, we acknowledge that to show that SPTs are generally a more robust alternative to estimating extinction rates, a systematic and widespread  analysis of the effect of incomplete sampling and detection on SPTs needs to be undertaken. This is still lacking and will be tackled in details in future works.

\subsection{Relations among SPT, SAR and EAR}

The SPT distribution is also particularly interesting because
it generates spatial patterns that can be related, within our theoretical framework, to other important macroecological patterns, as the SAR and the EAR relationships.
We can use the BSS data at
different spatial scales to investigate how SPTs can give
information also on the biogeography of the BSS plant ecosystem.
In fact, the scale-invariant character of $p_{\tau}$ indicates
that $\alpha$ depends only on the spatial connectivity of the
environment (i.e. networked, $2D,..$), and not on the sampling
area. On the contrary $\nu$, which accounts for immigration
processes from species outside the local community, is argued to
decrease with increasing sampling area, thus representing a clear
signature of the geographic scale of the analysis. The SAR
characterizes how the average number of observed species increases
with increasing sample area and it is usually characterized by a
power-law behavior, i.e. $S\sim A^z$ \citep{brown1995}. Our
theoretical framework allows for the linkage of the cut-off
timescale $1/\nu$ with the spatial scale of analysis, and the SPT
distribution to the SAR. In fact, in the neutral model, species
emerge as a point Poisson process with rate $\nu N$ and last for a
lifetime $\tau$ . Therefore, the mean number of species in the
system at stationarity is $\langle S \rangle=\nu N \langle \tau
\rangle$ \citep{bertuzzo2011}. By deriving the scaling behavior of $\nu$,
$N$ and $\langle \tau \rangle$ with respect to $A$, we can thus
make a connection between SPTs and the SAR.

The last term in Eq. (\ref{pL_final}) represents the analytical
expression of the fraction of species that are present during the
whole observational time period ($\mathcal{S}(\Delta t_w)$). It
provides a tool to quantitatively estimate the diversification
rate from the empirical SPT pdf. In fact, in order to infer with
better precision $\nu$ from the observational data, we numerically
find the value of $\nu$ that minimizes $\mathcal{S}(\Delta t_w)$
at every different spatial scales $A$ (see Appendix C for details)
using the scaling exponent $\bar{\alpha}$. The best fit of the
scaling law between $\nu$ and $A$ obtained in this way gives
$\nu\sim A^{-\beta}$, with $\beta=1\pm0.1$ (Figure 5a), thus
confirming that the diversification time $\nu^{-1}$ scales almost
linearly with the area \citep{Zillio2005}.

Once the scaling law between $\nu$ and $A$ is established, in
order to derive the SAR, we need to estimate the scaling between
the total number of individuals $N$ and the sampled area $A$. In
\citet{bertuzzo2011} an isometric relation $N \propto  A$
\citep{machartur1967} was assumed. However the abundance data on
the New Jersey plant communities show that this is not the case
for that ecosystem. In fact the average total species coverage
percentage over all the plots is $165\%$. Samplers start with the
topmost layer, usually the trees, and work their way down to the
species closest to the ground \citep{BSS}. Thus the exceedance in
the coverage percentage is due to the overhanging among species.
In order to evaluate the scaling relation between $N$ and $A$, we
then build a minimalist model, in which species are classified by
their typical average size. Assuming biotic saturation, i.e. that
each species covers completely the sampling region, in a plot of
area $A$ we can find at most $[A/a]$ individuals of characteristic
size area $a\ll A$ ($[x]$ denotes the smallest integer not greater
that $x$), $[A/2a]$ individuals of characteristic area $2a$ and so
forth. In general, the total number of individuals $N$ in a region
of area $A$ scales as
\begin{equation}\label{NA}
    N(A) \propto \sum_{n=1}^{[A/a]}\frac{A}{n a}\sim\frac{A}{a}\,H_{A/a},
\end{equation}
where $H_q=\sum_{n=1}^{q}1/n$ is the harmonic number and
$H_{A/a}\sim \ln(A/a)$. The constant of proportionality in Eq.
(\ref{NA}) depends on the arbitrary choice of $a$. However since
$H_{A/a}\sim \ln(A/a)+\,$sub-leading terms,  a change of $a$ in
$\lambda a$ leads to $H_{\lambda a} =
\frac{1}{\lambda}H_{A/a}+\,$sub-leading terms and so its precise
value is irrelevant since it can be absorbed in the
proportionality constant (see below) to be determined from the
best fit of the data.

Finally, we calculate the asymptotic behavior of the mean SPT
$\langle\tau\rangle = \int t p_\tau (t)dt$ (see Appendix C for
mathematical details) and we get:
\begin{equation}\label{mean_tau}
  \langle \tau \rangle \sim  \left\{
  \begin{array}{ll}
    A^{-\beta(\alpha-2)} & \hbox{for $\alpha<2$;} \\
    \ln(A) & \hbox{for $\alpha=2$.}
  \end{array}
\right.
\end{equation}

Putting all the above scaling results together, we find that for
the BSS plant ecosystem (which is characterized by $\alpha \approx 2$) the mean-field neutral
framework predicts a power-law SAR with logarithmic correction:
\begin{equation}\label{SAR}
\langle S(A) \rangle= K A^{1-\beta}\ln(A)\,H_A,
\end{equation}
where $K$ is the constant of proportionality. We determine the
constant $K$ in three different ways: 1) through the best fit of
Eq. (\ref{SAR}) on the empirical SAR data; 2) by imposing that in
the total area $A^*$ we have the total number of species $S^*$; 3)
by finding the value of $K$ that gives the exact number of species
in the smallest area ($A=96$ m$^2$) and then predicting the
complete SAR curve (upscaling). Remarkably, all three methods
yield $K\approx 3.46$. We also performed the best fit of the
power-law SAR $\big(S\sim A^z\big)$ on the empirical data,
obtaining $z=0.34$. We note that in our range of areas, this
result it is not distinguishable from the SAR given by Eq.
(\ref{SAR}) as shown by Figure 5c.

In the special case of mean field interactions, we can also
calculate the EAR. EAR is defined as the relation between the
number of species endemic to a region and the area of that region
\citep{Kinzig2000,green2003}. The EAR has been recently proposed
by \cite{he2011} as the correct approach in order to estimate
extinction rates from habitat loss. The fact that the mean-field
approach well describes the empirical SPT distribution suggests
that in the BSS ecosystem the phenomenon of spatial aggregation or
clustering is negligible. In fact mean-field dynamics indicates
that species composition is well mixed in the ecosystem and thus
it does not depend on spatial features. Therefore, in this case,
we can evaluate the EAR curve just by reversing the SAR
\citep{he2011}:
\begin{equation}\label{EAR}
   \langle E(A) \rangle =S^*-\langle S(A^*-A)\rangle= S^*-\ln(A^*-A)\,H_{A^*-A},
\end{equation}
where $E$ is the number of endemic species. Figure 5 summarizes
all the above results and shows the comparison between the
empirical and theoretical results for the SAR and the  EAR.

\section{Conclusions}

We have presented new theoretical and empirical evidence
supporting the broad validity of a recently proposed
macroecological pattern, the SPTs
distribution. Specifically, we have completed (and described in
detail) the theoretical treatment of the relevant ecosystem
dynamics and of the derived sampling problem, and analyzed
empirically two further rather diverse ecosystems (hosting
respectively herbaceous plants and estuarine fishes). In both cases
the observed SPT distributions display a power-law shape with a
cut-off due to the finiteness of the observational time window,
which is reproduced by the theoretical model. These results are
expected to be robust with respect to relaxations of specific
assumptions on ecological neutrality \citep{bertuzzo2011}, thus
confirming our expectation that power laws, observed for SPT pdfs,
are the result of emergent behaviors independent of fine details
of the system dynamics.

The specific values of the scaling exponents of the SPT distributions depend only on a
few key factors, namely the spatial dimension of the embedding
ecosystem matrix and the nature of dispersal limitation. In
particular, our results show that dispersal limitation does not
occur in the ecological dynamics of both ecosystems under study.
While this result could be expected for the estuarine fishr inventory,
where data were collected at a single point in space, quite
surprisingly the BSS plant community does not show signs of
dispersal-limited spatial interactions. We interpret this result
as a consequence of the fact that the total sampled area in the
BSS forest is very small. This suggests that the
average plant dispersal radius is comparable to the characteristic
size of the ecosystem, thus complying to mean-field-like dynamics.

Because of the assessed, robust scale invariance character of the SPT distribution (limited only by biogeographical
finite-size effects), and because of its relation with other macro-ecological patterns, we have proposed that the SPTs distribution is a powerful synthetic descriptor of ecosystem biodiversity and of its associated dynamics.

New data would allow for the investigation of other implications
of the theoretical predictions. For instance, the neutral
voter model predicts that the scaling exponents of the SPT
distributions of riparian ecosystems (i.e. networked environments
where directional, anisotropic dispersal is forced by the
structure of the fluvial environmental matrix) should be lower
than those of 2D, 3D or mean-field ones. This result, if proven,
would have remarkable consequences for conservation ecology,
because it suggests that species that disperse isotropically have
shorter average persistence times than species that are
constrained to disperse along spatially constrained and
anisotropic ecological corridors, like those provided by river
networks. This, in turn, calls for long-term analysis of riparian
ecosystems to test empirically the effects of river morphology on ecosystem dynamics.

For these reasons we suggest that field
biologists and ecologists should perhaps invest renewed efforts in
collecting improved long-term datasets of ecosystem dynamics at
different spatial and temporal scales and under different
dispersal conditions (for instance, archives beyond
presence/absence of species within networked environments like
riverine ecosystems for freshwater fish) as they prove crucial for
our deeper understanding of universal patterns in macroecology.
The present study shows that long-term datasets can be profitably
used to highlight crucial properties and spatial effects of
ecosystem dynamics, including the role played by the underlying
connectivity structure in shaping SPT distributions.

\flushleft{\emph{Acknowledgments}}\\
We are grateful to Pisces Conservation Conservation and to Dr. P.
Henderson for granting us open access to the Hinkley fish
database. We acknowledge the BSS research group for making their
data available through the NSF grant DEB 97-26992 (Long-Term
Research in Environmental Biology). We  acknowledge two reviewers for their suggestions and comments 
that  have been instrumental in our revision, and have remarkably improved the
manuscript.   SS, EB, LM and AR gratefully
acknowledge the support provided by ERC advanced grant program
through the project RINEC-227612 and by the SFN/FNS project
$200021\_124930/1$. IRI gratefully acknowledges the support of the
James S. McDonnell Foundation through a grant for Studying Complex
Systems (220020138). AM acknowledges funds provided by Fondazione
Cariparo (Padova, IT).

\appendix


\section{}
In this appendix we derive the analytical results of the
stick-length sample problem presented in the main text.
The set up of the thought experiment is fully described in the main text and in Figure 2.

Specifically, the stick-length sample problem can be characterized by five random variables, namely:
i) $X^0$ is the position of the left side of a given stick; ii) $L$ describes the
length of a stick; iii)  $L_c$  is the length of those sticks that
cross $x^*$; iv)  $L_p$ is the length of that part of the stick that
crosses $x^*$ or $\hat{x}$ and fall within $\Delta x$; v) $L'$ is the length of a stick that is totally or
partially comprised inside $\Delta x$. 

{$X^0$ and $L$ are the two independent random variables of the problem, while $L_c$, $L_p$ and $L'$ can all be expressed in terms of $X^0$ and $L$, as we will show in details. Assuming that $p_{L}$ is the unbiased length stick distribution, we now present a
methodology that enable us to derive the length pdf $p_{L_c}$, and
the pdf $p_{L_{p}}$ as a function of $p_{L}$. The same methodology
may be applied to obtain $p_{L'}$

The random variable $L_p$ can be expressed as a function
of the random variables $L$ and $X^0$. In fact, we have that the
sticks that cross $x^*$ are those for which $X^0<x^*$ and
$X^0+L>x^*$  (see Figure 2). Therefore, if $\Theta$ is the
Heaviside step function, the length of a stick that crosses $x^*$
is given by the random variable
\begin{equation}\label{tau_c}
   L_c= L \Theta(X^0+L-x^*)\Theta(x^*-X^0).
\end{equation}
We know that $L$ is distributed ($\sim$) as $p_L$ and that $X^0$ is uniformly distributed along the $x$-axis (in particular $X^0\sim U_{(0,x^*)}$). Therefore we have that the probability density function (pdf) $p_{L_c}$ is
\begin{equation}\label{p_tau_c1A}
    p_{L_c}(l)=\frac{\langle \delta(l-L)\Theta(X^0+L-x^*)\Theta(x^*-X^0)\rangle}{\langle\Theta(X^0+L-x^*)\Theta(x^*-X^0)\rangle},
\end{equation}
where $\delta$ is the Dirac delta function. The numerator of Eq. (\ref{p_tau_c1A}) simply gives the ensemble average, taken over the random variables $X^0$ and $L$, of different realizations of the event that occurs when a stick of length $L$ intersects $x^*$, while the denominator in (\ref{p_tau_c1A}) assures the normalization condition.

In order to solve the ensemble average of the latter equation, we split the calculation into two steps. First we calculate the conditional probability function $p_{L_c}(l|L)$ of a stick crossing the vertical line $x^*$, given the fact that it has a length $L$, i.e. we perform only the average over $X^0$
\begin{equation}\label{p_tau_c|tau}
    p_{L_c}(l|L)=\frac{\int_{\max[0,x^{*}-L]}^{x^*}\delta(l-L)dX^0}{\langle\Theta(X^0+L-x^*)\Theta(x^*-X^0)\rangle}.
\end{equation}

As a second step, we marginalize Eq. (\ref{p_tau_c|tau}) over $L$, i.e. we perform the average also over $L$:
\begin{equation}\label{p_tau_c}
    p_{L_c}(l)=\frac{\int_0^{+\infty}p_{L}(L)\delta(l-L)\min[x^*,L]dL}{\int_0^{+\infty}\min[x^*,L]p_{L}(L)dL}.
\end{equation}

Assuming that we do not have boundary effects (i.e. $x^* \rightarrow+\infty$) we obtain Eq. (\ref{p_L_c}) in the main text.

The case of $L_{p}$ can again be obtained using the same technique. Writing the random variable $L_{p}$ as a function of $L$ and $X^0$ (see figure 2) we get
\begin{equation}\label{tau_cA}
    L_{p}=(x^*-X^0)\Theta(x^*-X^0)\Theta(X^0+L-x^*)\Theta(X^0).
\end{equation} Following the same steps presented before, we obtain
\begin{equation}\label{p_tau_m|tau_cA_0}
    p_{L_p}(l|L)=\frac{\int_{0}^{\infty}\delta(l-(x^*-X^0))\Theta(x^*-X^0)\Theta(X^0+L-x^*)\Theta(X^0)}{\langle\Theta(x^*-X^0)\Theta(X^0+L_p-x^*)\rangle},
\end{equation}
and marginalizing over $L$ we obtain

\begin{equation}\label{p_tau_m|tau_cA}
    p_{L_p}(l)=\Theta(x^*-l)\frac{\int_{0}^{\infty}\Theta(L-l)p_{L}(L)}{\int_0^{+\infty}\min[x^*,L]p_{L}(L)dL},
\end{equation}

that, in the limit $x^*\rightarrow\infty$, results in Eq. (\ref{p_L_m_sol}).
The case of a stick crossing $x=\hat{x}$ is perfectly symmetric to the one discussed above and would lead to the same result.

The results for $L'$ and $L_I$ follow similarly.

\section{}
In this section we present the calculations used to estimate the diversification rate $\nu$ from the empirical SPT distributions of the BSS plant community.
We denote by $\mathcal{E}(\Delta t_w)$ the fraction of species, evaluated from the data, that are present during the whole observational time period $\Delta t_w$.
The last term in Eq. (\ref{pL_final}), denoted here as $\mathcal{A}_{\Delta t_w}$, gives the theoretical value of the latter quantity, i.e. the atom probability of the SPT distribution for $t=\Delta t_w$. By using the theoretical SPT distribution described by Eq. (\ref{SPT_distribution}) we can estimate $\mathcal{A}(\Delta t_w)$:

\begin{eqnarray}
\nonumber \mathcal{A}_{\Delta t_w}(\nu) &=&\int_{\Delta t_w}^{\infty}(\tau-\Delta t_w) \frac{\nu ^{1-\alpha}}{\Gamma(1-\alpha,\nu t_{min})} {\tau^{-\alpha}e^{-\nu \tau}}d\tau\\
 \label{atomo} &=&\frac{\Gamma (2-\alpha ,\nu  \Delta t_w)-\nu  \Delta t_w \Gamma (1-\alpha ,\nu  \Delta t_w)}{\nu  \Gamma (1-\alpha ,\nu t_{min})},
\end{eqnarray}

where $\Gamma[a,b]$ is the incomplete Gamma function, and $t_{min}$ is the minimum SPT observed in the ecosystem (in general $t_{min}\rightarrow0$). If $\alpha<2$, then the dependence of Eq. (\ref{atomo}) on $\nu t_{min}$ is weak, and thus, as $\nu t_{min}\ll1$, we can set $\nu t_{min}\approx 0$.

Therefore we calculate the values of the diversification rate $\nu$ for a given $\Delta t_w$ (that in turn depends on the specific sample area $A$) by numerically minimizing the quantity

\begin{equation}\label{nuA}
   \mathcal{S}(\nu|\Delta t_w)= ( \mathcal{A}_{\Delta t_w}(\nu)-\mathcal{E}(\Delta t_w))^2,
\end{equation} where the average scaling exponent $\bar{\alpha}$ obtained by the best fit of the SPT pdf $p_{\tau'}$ and $p_{\tau_I}$ has been used for $\alpha$ .
Eventually, repeating the same procedure using different time windows (corresponding to the different sampled areas) we obtain the scaling relation between $\nu$ and $A$ presented in the main text.
\section{}
In this appendix we derive the asymptotic behavior of the mean persistence times $\langle \tau \rangle$. We divide the calculation into two cases: $\alpha<2$, where the dispersal limitation is an important driver of the ecosystem dynamics and $\alpha=2$, the mean field case.
The mean persistence time for $\alpha<2$ is
\begin{equation}\label{mean_Persistence time_A}
    \langle \tau\rangle=\frac{\int_{\nu t_{min}}^{\infty}t^{1-\alpha}e^{-t}dt}{\nu\int_{\nu t_{min}}^{\infty}t^{-\alpha}e^{-t}dt}=\frac{\Gamma(2-\alpha,\nu t_{min})}{\nu\Gamma(1-\alpha,\nu t_{min})}.
\end{equation}
Integrating by parts, the denominator of Eq. (\ref{mean_Persistence time_A}) becomes:
\begin{equation}\label{sv_as_gamma}
 \Gamma(1-\alpha,\nu t_{min} )=\frac{(\nu t_{min}  )^{1-\alpha }}{\alpha-1}+\frac{1}{1-\alpha}\Gamma(2-\alpha,\nu t_{min} ),
\end{equation}
and using the limit $\lim_{\nu t_{min}\rightarrow0}\Gamma(2-\alpha,\nu t_{min} )=\Gamma(2-\alpha)>0$, Eq. (\ref{mean_Persistence time_A}) simplifies to
\begin{equation}\label{mean_Persistence time_A_final}
    \langle \tau\rangle\sim\nu^{\alpha-2}.
\end{equation}
In the case $\alpha=2$ we have instead
\begin{equation}\label{mean_Persistence time2A}
    \langle \tau\rangle=\frac{t_{min} E_1(t_{min} \nu ))}{E_2(t_{min} \nu )},
\end{equation}
where $E_n(z)=\int_1^{\infty}e^{-z t}t^{-n}dt$ is the exponential integral function.
The expansion of the exponential integral function for $x\approx 0$  is \citep[][pag. 229]{abramowitz1965}:
\begin{eqnarray}\label{sv_as_gamma}
E_1(x)&\sim&-\frac{x^2}{4}+x-\log (x)-\gamma\\
E_2(x)&\sim&-\frac{x^2}{2}+x (\log (x)+\gamma -1)+1,
\end{eqnarray}
where $\gamma$ is the Euler constant.
Substituting the latter expansion into Eq. (\ref{mean_Persistence time2A}) we obtain
\begin{equation}\label{mean_Persistence time3}
    \langle \tau\rangle\sim - t_{min }(\log (\nu )+ \gamma )\sim-\log (\nu).
\end{equation}
Therefore we find that $\langle \tau\rangle$ scales as the logarithm of $\nu$ and, in turn, as the logarithm of the sampled area $A$ (in fact we found that $\nu\sim A^{-1}$).
The logarithmic correction in the scaling behavior of $\langle \tau\rangle$ is supported by the data (see Figure 5 in the main text) confirming the hypothesis that the evolution of the BSS ecosystem is well mimicked by mean field dynamics.


\bibliographystyle{elsarticle-harv}



\clearpage
\newpage

\begin{figure}[h!]
  \includegraphics[width=25pc]{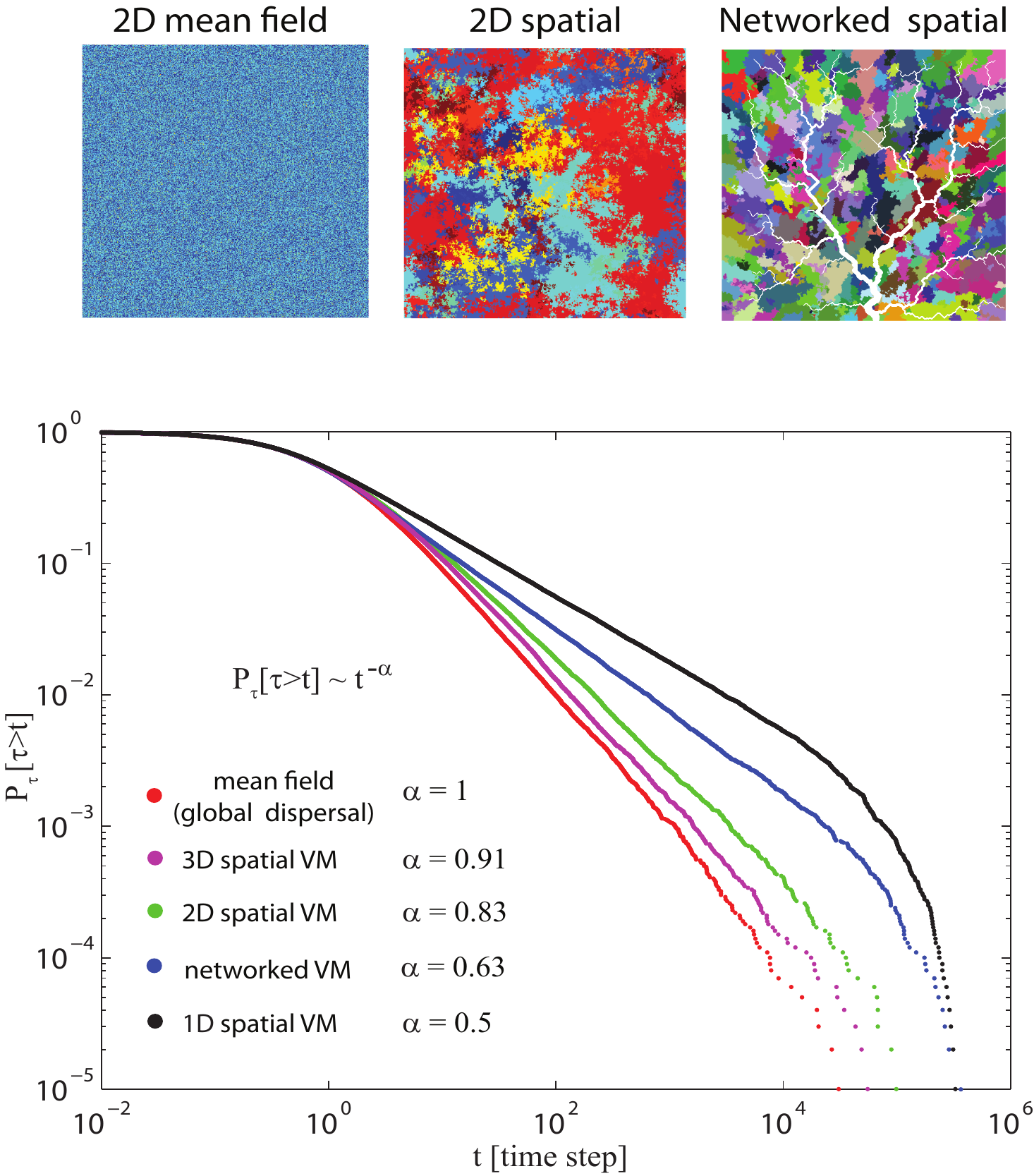}\\
  \caption{ (a) Persistence-time
exceedance probabilities $P_\tau(t)$ (probability that species'
persistence times $\tau$ be $\geq t$) for the neutral
individual-based model with nearest-neighbor dispersal
implemented on the different topologies (redrawn from \cite{bertuzzo2011}). Species persistence time distributions are crucially dependent on the type of spatial connectivity interactions in the voter model. }\label{figure1}
\end{figure}

\clearpage
\begin{figure}[h!]
  \includegraphics[width=19pc]{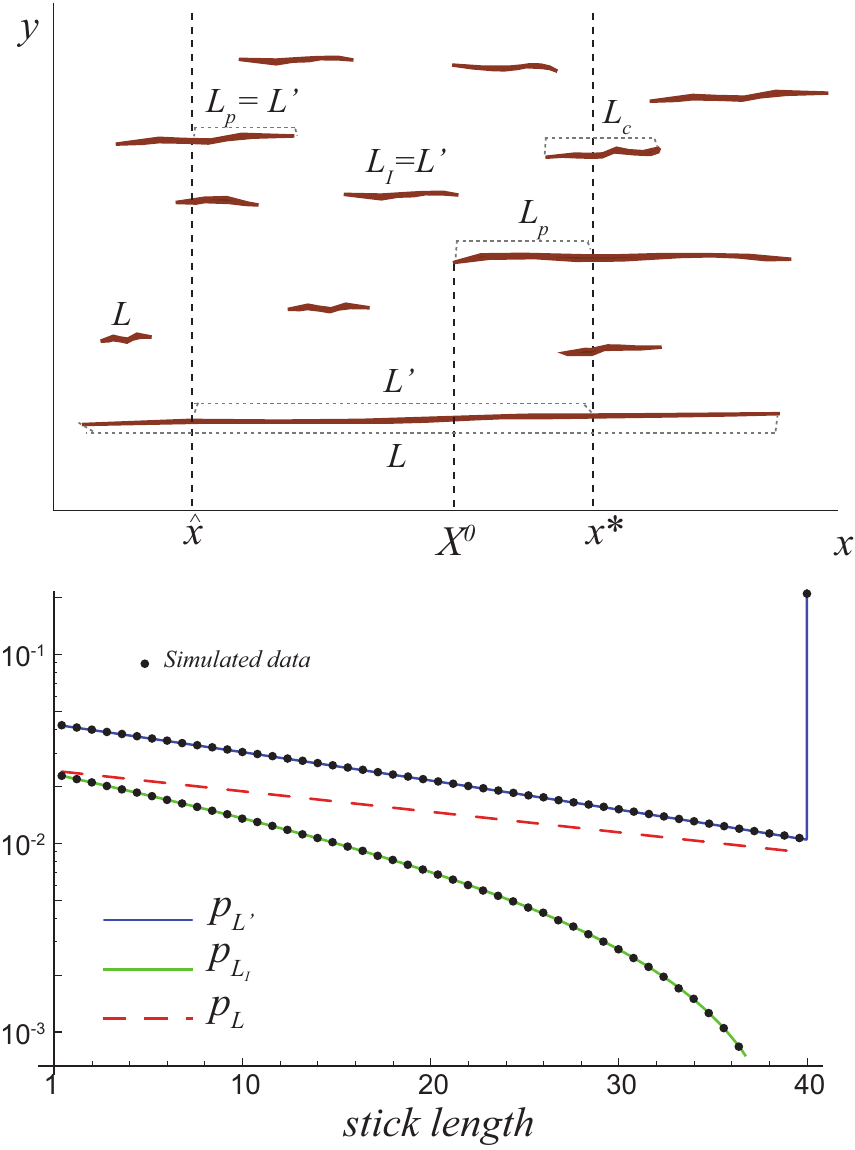}\\
  \caption{Top panel: schematic representation of the stick-length sample problem. The variables that can be measured from the sample of sticks in $\Delta x$ are $L'$ and $L_I,$ both with a different distribution with respect to $p(L)$. Bottom panel: comparison between analytical SPT  pdfs $p_{L'}$ and $p_{L_I}$ given by Eqs (\ref{pL_final})-(\ref{pL_i}),  and those obtained through numerical simulation of the stick length problem starting from an unbiased stick length distribution $p_L(l)=\lambda e^{-\lambda l}$, with $\Delta x=1/\lambda$  (in this particular example $\lambda=0.025)$. $p_L$, $p_{L_I}$ and $p_{L'}$ have been shifted in the semi-log $y$ plot for clarity.}\label{figure2}
\end{figure}

\clearpage
\begin{figure}[h!]
  \includegraphics[width=19pc]{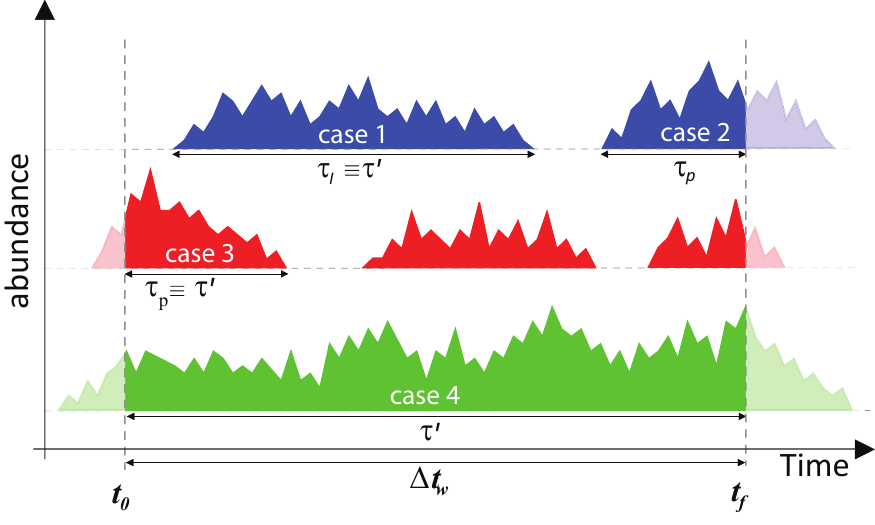}\\
  \caption{The species persistence-time (SPT) within a geographic region is defined as the time incurred between a species' emergence and its local extinction. Recurrent colonizations of a species define different SPTs
times. The SPTs $\tau_p$,  $\tau_I$ and $\tau'$, that can be measured within the observational window, are in general different from the unbiased SPT $\tau$. The four different cases are discussed in the main text.}\label{figure3}
\end{figure}

\clearpage
\begin{figure}[h!]
  \includegraphics[width=29pc]{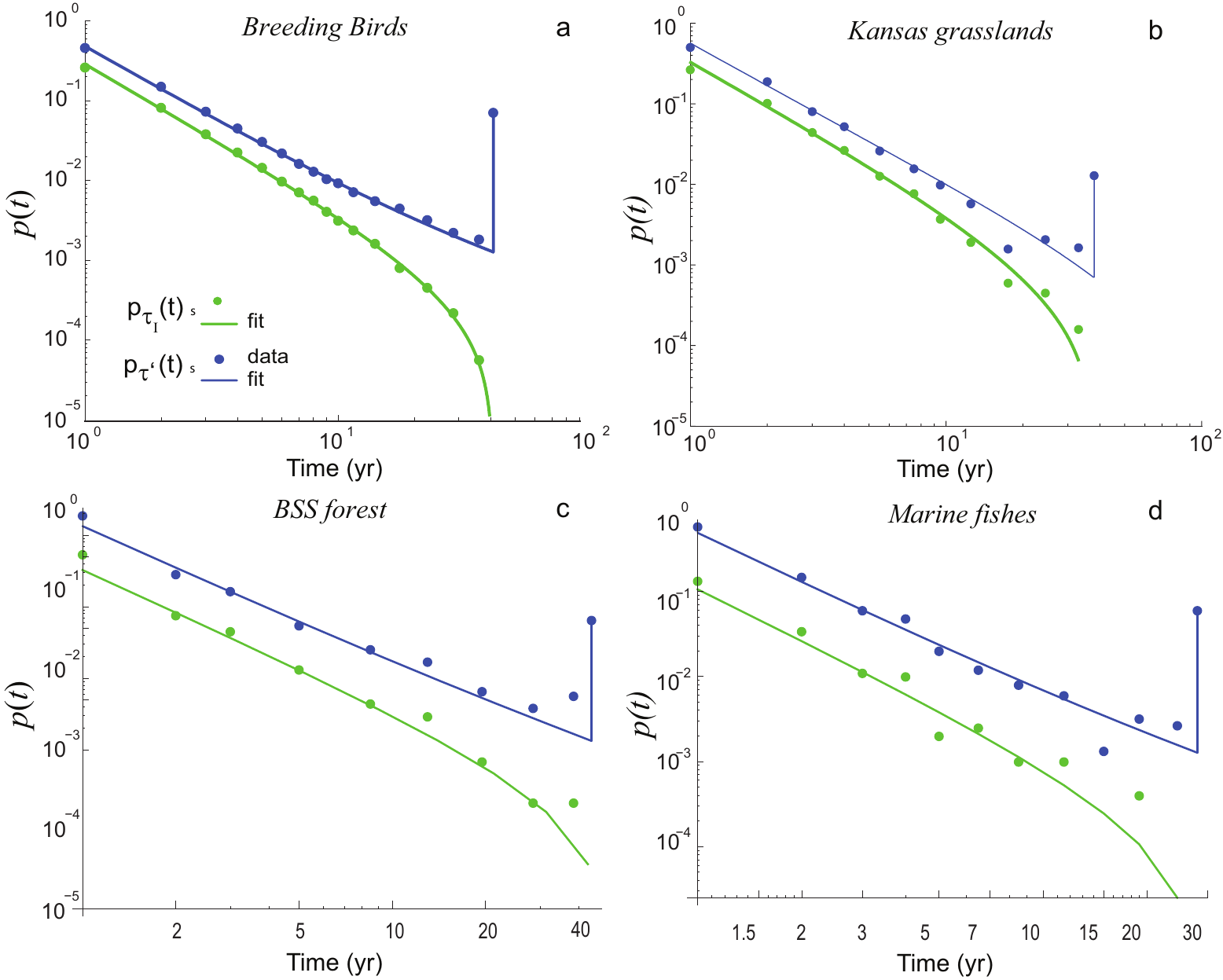}\\
  \caption{Comparison between empirical distributions for (a) North American Breeding birds, (b) Kansas grasslands, (c) New Jersey BSS forest, (d) estuarine fish community  and  the corresponding theoretical SPT pdfs $p(t)$ of $\tau'$ (green), $\tau''$ (blue). Filled circles and solid lines show
observational distributions and fits, respectively. The spatial scale of
analysis is $A= 10,000$ km$^2$ and $\Delta t_w=41$ years for (a),
A=1 m$^2$ and $\Delta t_w=38$ years for (b), A=480 m$^2$ and $\Delta t_w=44$ years for (c) and $\Delta t_w=28$ years for (d) (for details on the analysis of the databases relevant to (a) and (b) see supplementary material in \citealp{bertuzzo2011}). The finiteness of the time
window imposes a cut-off to $p_{\tau'}(t)$ and  an atom of probability in $t=\Delta t_w$ to $p_{\tau''}(t)$, which corresponds to the fraction of species that are always present during the observational time. $p_{\tau_I}(t)$ and $p_{\tau'}(t)$ have been shifted in the log-log plot for clarity.}\label{figure4}
\end{figure}
\clearpage
\begin{figure}[htcb]
  \includegraphics[width=19pc]{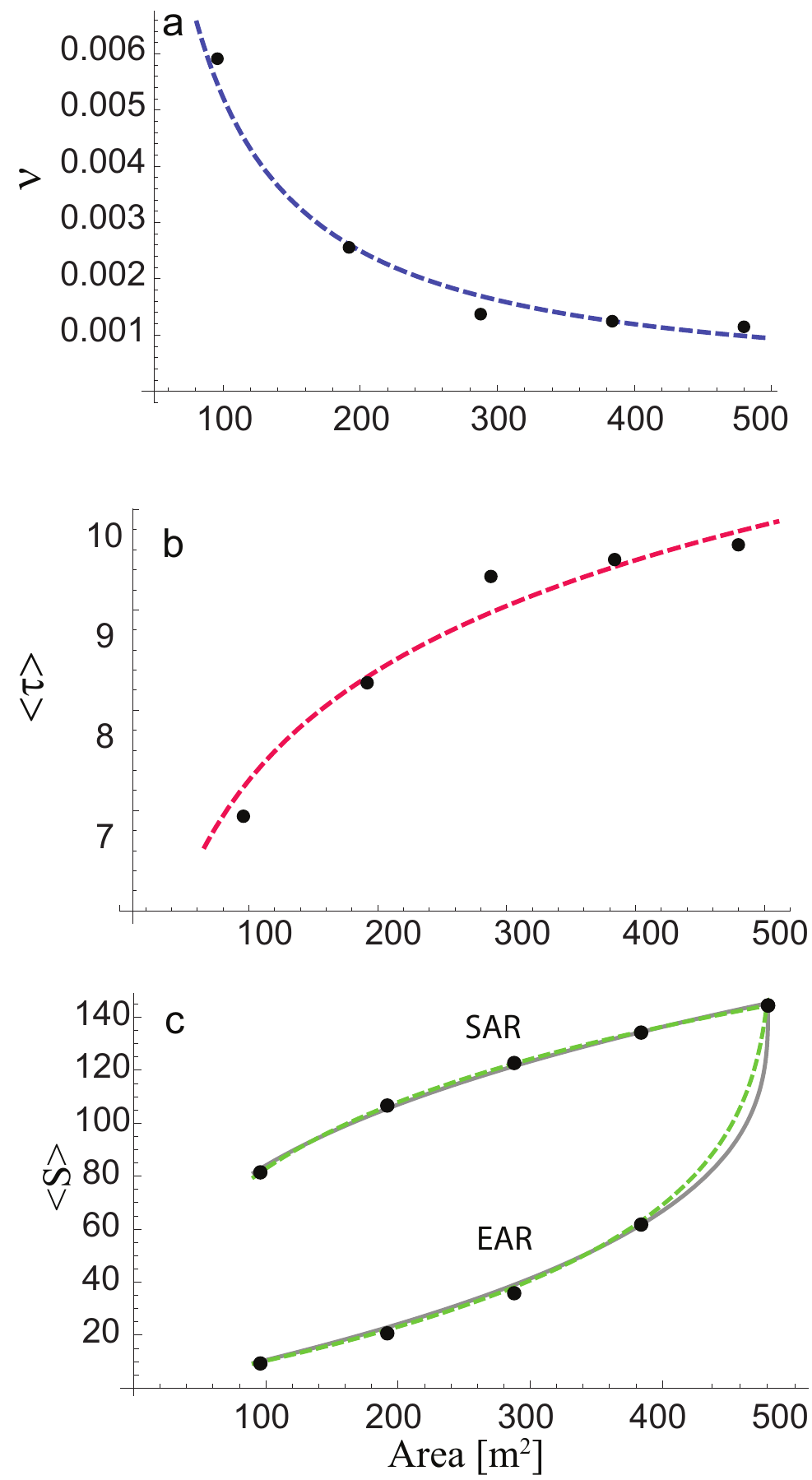}\\
  \caption{Comparison between a) diversification rates for different areas (black dots) and the scaling law $\nu \sim A^{-1}$ (dashed blue line);
b) mean persistence time for each sample area $\langle \tau(\nu(A),\bar{\alpha})\rangle$(black dots) calculated using Eq. (\ref{SPT_distribution}), and logarithmic relation $\tau\sim \ln(A)$ (dashed red line) predicted by the neutral model for the mean field case ($\alpha=2$);
c) empirical SAR and EAR relationships (black points), power law fitting of the SAR and the corresponding EAR (gray continuous line) and the mean field SAR and EAR given by Eqs. (\ref{SAR}) and (\ref{EAR}) (green dashed).}\label{figure5}
\end{figure}

\clearpage
\begin{table}
  \centering
  \begin{tabular}{lccccc}
    \hline
  \textbf{ecosystem}& \textbf{A} \textbf{[m$^2$]} &$\mathbf{\alpha}$ & $\mathbf{\nu}$\\\hline
    BSS plants & 98 & 1.97 & 0.0057 \\
    BSS plants & 196 & 1.97 & 0.0024 \\
    BSS plants & 294 & 1.96 & 0.0013 \\
    BSS plants & 382 & 1.96 & 0.0011\\
    BSS plants & 480& 1.96 & 0.0010 \\ \hline
    Hinkley fish & / & $1.97\pm0.06$ & $0.00049\pm0.00007$  \\ \hline
  \end{tabular}
  \caption{Exponents for the Hinkley Point estuarine fish database and for the BSS dataset at every spatial scale of analysis. The values of the exponents are obtained by the simultaneous best fitting of Eqs. (\ref{pL_final}) and (\ref{pL_i}) on the empirical SPT distributions. They suggest that both ecosystems are driven by a mean field dynamics. The value of $\bar{\alpha}$ given in the text corresponds to $\bar{\alpha}=(\sum_{i=1}^{5}\alpha_i)/5=1.97\pm \sigma_{\bar{\alpha}}$, where $\sigma^2_{\bar{\alpha}}=\sum_{i=1}^5\sigma^2_{\alpha_i}/5=0.015$ ($\sigma^2_{\alpha_i}$ is the variance for the $i-$th $\alpha$ exponent from the fit). Values of $\nu$ for the BSS plants are obtained by the minimization procedure explained in the text and in Appendix B. }\label{fit_result}
\end{table}
\clearpage

\end{document}